\documentclass{aip-cp}
\usepackage[numbers]{natbib}
\usepackage{rotating}
\usepackage{graphicx}

\usepackage{hyperref}
\usepackage{color}
\usepackage{epstopdf}
\usepackage{cleveref}
\usepackage[svgnames]{xcolor}
\usepackage{enumerate}
\usepackage{braket}
\hypersetup{hidelinks,colorlinks=true,allcolors=DarkBlue}

\begin{document}

\title{Practical Cryptographic Strategies in the Post-Quantum Era}

\author[aff1]{I.S. Kabanov}
\author[aff2]{R.R. Yunusov}
\author[aff2]{Y.V. Kurochkin}
\author[aff2]{A.K. Fedorov}\corresp[cor1]{Corresponding author: akf@rqc.ru (A.K.F.)}

\affil[aff1]{Schneider Electric, Cambridge, Massachusetts 02142, USA.}
\affil[aff2]{Russian Quantum Center, Skolkovo, Moscow 143025, Russia}

\maketitle

\begin{abstract}
Quantum key distribution technologies promise information-theoretic security and are currently being deployed in commercial applications.
We review new frontiers in information security technologies in communications and distributed storage applications with the use of classical, 
quantum, hybrid classical-quantum, and post-quantum cryptography. 
We analyze the current state-of-the-art, critical characteristics, development trends, and limitations of these techniques for application in enterprise information protection systems. 
An approach concerning the selection of practical encryption technologies for enterprises with branched communication networks is discussed. 
\end{abstract}

\section{Introduction}

Universal quantum computers would allow solving certain mathematical problems in a more efficient way in compare with their classical counterparts. 
In particular, tasks dealing with integer factorization and discrete logarithm problems~\cite{Shor1997}.
Therefore, any security protocols, products or security systems~\cite{Schneier1996} that derive security from mathematical complexities of the above tasks are highly vulnerable 
to attacks with the use of quantum computer~\cite{ETSI}.
On the one hand, a question about the time, which is needed to build of a large-scale quantum computer, remains open. 
There are also a number of physical setups that realize different implementations of quantum computing devices such as superconducting qubits, ultracold atoms, and trapped ions~\cite{Lukin2016}. 
On the other hand, the fact of the possible appearance of quantum computer in the near future changes development trends of information security. 

Examples of this paradigmatic change include an increase of the attention to security tools that are not vulnerable to attacks with the use of quantum computer, 
so-called quantum-safe technologies. 
Generally speaking, these techniques can be divided into two classes. 
The first is based on information-theoretically secure methods since they make no assumptions about computational resources of an eavesdropper, 
e.g., one-time pad~\cite{Shannon1948} and Wegman-Carter authentication~\cite{Wegman1981} schemes. 
However, the one-time pad scheme requires the establishing secret keys. 
Remarkably, the procedure of key distribution can be realized in an efficient manner using individual quantum systems, single photons~\cite{Gisin2002}. 
Recent breakthrough in experimental control for quantum systems aimed 
on a design of quantum key distribution (QKD) devices for practical usage of the one-time pad scheme has been performed~\cite{Gisin2002}. 
QKD systems combining with other cryptographic tools are used in the hybrid classical-quantum encryption systems~\cite{IDQ}. 
Cryptographic devices with QKD are available on the market~\cite{IDQ}. 

The second class of tools is based on computational problems that are currently believed to be hard both for classical and quantum computing devices.
In particular, one can use post-quantum cryptography primitives~\cite{Bernstein2009} or symmetric ciphers like AES (with the use of a quantum-resistant key distribution scheme),
since the cipher can adapt to a quantum attack by increasing its key size to rectify a vulnerability introduced by quantum computing.

The main goal of the paper is to analyze critical characteristics, trends, and limitations of these technologies. 
We stress on application of practical cryptographic tools in enterprise security systems. 
It is clear that typically such security systems deal with information of varying degrees of importance. 
Therefore, it is useful to consider selecting, which proper cryptographic techniques aimed on different applications such as secure communications and distributed storage. 
Due to increasing of possibilities for data storage during last decades, 
analysis of development trends is especially important for long-term security in the view of ``store now --- decrypt later'' paradigm,
in which we assume that sensitive data can be stored now in an encrypted form and then decrypted when quantum computers and/or novel mathematical algorithms are available. 

\section{Cryptographic tools: quantum, post-quantum, and hybrid}

Since not all cryptographic algorithms are vulnerable to quantum attacks, 
there is an increase of the interest to security methods providing long-term information protection robust against attacks from quantum computers. 
As it was mentioned below, several ways to provide information security using cryptography in the view of appearance of a universal quantum computer can be used. 

\subsubsection{Quantum key distribution for information-theoretic secure systems}

We start with a description of the use of secret key cryptography in the one-time pad regime with QKD. 
The operating of such a method can be described as follows. 
Two legitimate users (Alice and Bob) have the pre-shared authentication key and the direct transmitting channel, i.e. Alice and Bob should be point-to-point connected to each other. 
Then they establish a QKD session that allows them to obtain a raw quantum key, which contains some errors. 
In the QKD security proofs it is assumed that all errors in raw quantum keys are due to eavesdropping~\cite{Gisin2002}. 
That is why Alice and Bob use authenticated public channel for the post-processing procedure~\cite{KiktenkoFedorov2016,KiktenkoFedorov2017}. 
After the procedure, Alice and Bob have a key for applications, and it is proven to be information theoretically secure against arbitrary attacks, including the quantum ones~\cite{Scarani2009}. 

We note that the practical usage of the one-time scheme with QKD encounters a number of practical challenges. 
First, it is an essential limitation that Alice and Bob should have a direct (fiber or free-space) channel for transmitting of single photons and authenticated classical channel for information reconciliation. 
Second one is that the key generation rate is rather low. 
It decreases significantly with increasing distance between Alice and Bob due to the optical losses of photons in optical fibers and the imperfect work of single photon detectors. 
Towards to overcome this challenge one needs to develop new generation of single photon sources and detectors. 
Finally, the post-processing procedure for information reconciliation includes computationally cost operations, e.g., error correction with low-density parity check codes~\cite{KiktenkoFedorov2017}. 
Nevertheless, elusively the one-time pad scheme with QKD is both practical and absolutely secure tool. 
This means that even assuming that Eve has unlimited computational resources, classical or quantum, QKD is always will be secure~\cite{Gisin2002,Scarani2009}. 

\subsubsection{Post-quantum cryptography}

The second way is to use post-quantum cryptography tools~\cite{Bernstein2009}. 
Such schemes are based on schemes, e.g. code-based, multivariate, lattice-based, and hash based cryptosystems, 
for those there are no both classical and quantum efficient algorithms.
Speaking about applications, post-quantum cryptography schemes have performance comparable to or better than pre-quantum schemes of the same security level. 
However, key, message, and signature sizes are generally larger. Furthermore, these schemes can be useful, however they do not guarantee absolute privacy. 
As well as classical public-key cryptography primitives these methods are fully non-resistant to ``store now --- decrypt later'' paradigm, 
because of an existence of the possibility to invent a  ``post-quantum computer''.

\subsubsection{Hybrid cryptography}

Lastly, a useful strategy is combination of different cryptographic techniques. 
For example, one can combine QKD with classical block ciphers (hybrid classical-quantum cryptography) with an increase of frequency of the master key update. 
This idea is used in commercial QKD devices~\cite{IDQ}. 
Several information security applications allow one also combine keys distributed using, let say, public-key cryptography and QKD. 
Hybrid systems such, as systems using QKD together with the AES algorithm, are quantum-safe. 

Let us consider a simple model of a hybrid classical-quantum system. 
In a classical part, one has a master key $K_M$, which is installed at starting point of the system, 
and session key $K_S$ obtaining with the use of non-quantum-safe tools. Alice and Bob use a function $d(K_M,K_S,M)$ to encrypt a message $M$. 
Using parameters of keys $K_M$ and $K_S$ one can estimate the time of information being secure $T_S$ in the view of possible attacks. 
A quantum part of hybrid systems consist of upgrade of the master key $K_M$ with a frequency $f$ by using a key $K_Q$  from quantum-safe key distribution scheme (e.g., QKD) as follows: 
$K_{M+1}=g(K_M,K_Q )$. 
Therefore, in the hybrid system one has $d(g(K_M,K_Q ),K_S,M)$ for encryption messages, 
and the time of information being secure $T_{S+Q}(f)>T_S$, and it is a function of the frequency $f$ of the master key upgrade. 
It is an important problem to find an algorithm, which allows one to obtain a good estimation for the time $T_{S+Q}(f)$. 

Another interesting idea is to combine QKD with classical approaches for authentication, e.g., as in the floating bases protocol for QKD~\cite{Kurochkin2005,Trushechkin2017}. 
Not only quantum key can be applied to the classical cryptography but also classical algorithms become stronger being moved to the quantum world. 
In the classical world the large computational power threats the transmitted data by the  ``store now --- decrypt later'' attack. 
In the quantum case even interception of the signal does not give all information because of the measurement properties. 
This approach was also used in quantum data transmission~\cite{Lloyd2016} and quantum authentication~\cite{Fehr2016}. 
The quantum authentication its one of directions where low bit rate of quantum channels can be applied for corporate needs.

\section{Cryptography in enterprise security systems}

We expect that enterprises information protection strategies will transform under influence of emerging quantum computers. 
There is no single universal standard for encrypting all data today and in the post-quantum era enterprises, 
governments, and public institutions will be challenged by similar tradeoffs while defining an effective strategy. 

Complete transition of all encryption system users to quantum resistant encryption will not happen in a moment, 
new standards and products resistant to quantum computing techniques will evolve along the time, 
but organizations today already should design their information protection tools considering the upcoming quantum resistant algorithm transition. 
Ref.~\cite{ETSI} gives an estimation for the time: 
"If a large-scale quantum computer (time -- $z$) is built before the infrastructure has been re-tooled to be quantum-safe and the required duration of information-security has passed (time -- $x+y$), 
then the encrypted information will not be secure, leaving it vulnerable to adversarial attack". 
It is then important to note that time estimate for making our IT infrastructure quantum-safe should include work on standardization of quantum-safe and hybrid encryption tools. 

A proper encryption strategy should depend on the sensitivity of organization's information, data storage and transmission methods. 
Before selecting adequate encryption tool organizations have to determine subjects of encryption and plan an encryption program 
as a part of an overall enterprise risk management and data governance program. 
We have in mind a simple model of an organization with a number $N$ of distinct branches. 

Selection of encryption technology is always made under various constraints, including legal, technological, financial, and organizational and a carefully planned, 
comprehensive approach that considers specifically which data should be encrypted and how will generate greater efficiency and effectiveness for the organization. 
We denote $C=\{c_1,c_2,\dots,c_M\}$ as a set classifying the information in an organization, where $c_1$ is a class of open (public) information and $c_M$ stands for the most critical information.  
We assume that the cost of information is significantly higher that the cost of implementation of quantum-safe tools. 

Another dimension that organization should take into account while planning their emerging encryption strategies is how the data will be protected throughout its lifecycle. 
It is therefore important to consider the state of the data under protection: 
data being transmitted over a network, data at rest, or data in the process of being generated, updated, erased, or viewed. 
Each of these states presents unique challenges and significantly influence what encryption technique is going to be used to secure data. 
We denote $T=\{t_1,t_2,…,t_K\}$ as a set of time classes of the information, where $t_1$ denotes the class of information, 
which should be private during only one time period of its lifecycle, and $t_K$ stands for the information class, which should be private during whole lifecycle.

Protecting data at rest is a critical issue as the network perimeter continues to dissolve. 
Major encryption types for data-at-rest include full disk encryption, hardware security module, encrypting file system, files and databases for protecting structured and unstructured data. 
Encryption types for data-in-motion include network protected access and server communications. 
Cloud computing has created the need to secure data in use as third-party providers increasingly host and process data. 
The data-in-use is the hardest to protect, since it almost always has to be decrypted and therefore exposed in order to be used. 
Specifically, this challenge is related to decryption keys, which must be completely unavailable 
to an attacker in order for encryption to provide security therefore protecting an environment where keys are stored is critical to address in an encryption strategy. 

It was mentioned previously 
that encryption should be a part of a broader security strategy of an organization and effective data classification is crucial for enabling resilient data protection capabilities for the organization. 
Data mapping exercise should done and take into account where information is stored ensuring that data in all locations such as mobile devices, 
backup systems and cloud services will be properly protected. 
Another challenge, which highlights the importance of data classification, 
is that appearance of a large-scale quantum computer may put at risk data encrypted by non-quantum resistant algorithms, which have significant long-term value along its lifecycle before retirement. 

On the basis of the analysis of critical characteristics, development trends, 
and limitations of these techniques for application in enterprise security systems we propose a framework for making practical trade off. 
The proposed framework gives trivial answers for the limit cases.
First, if data has the lowest class from the sets $C$ and $T$, then one can use the simplest and cheapest method for the protection. 
Second, if the considered information has the highest class from the sets $C$ and $T$, then one has to use quantum-safe systems such as QKD. 
In the other cases, one can use hybrid encryption methods with different parameters. 
It is possible to choose proper parameters of the used hybrid systems, such as the size of keys (master, quantum, and session) and the master key upgrade frequency, to realize practical trade offs. 

\section{Architecture of networks: Many-to-one hybrid systems}

Here we suggest combining of all different encryption techniques with taking into account their critical characteristics, 
development trends, and limitations of these technologies in special network architecture. 
It can be directly integrated in enterprise security systems. 

On the top of the considered above information characteristics, one should note a network topology of the considered organization. 
Using of quantum-safe security tools based on QKD require a direct channel communication channel. 
We begin with main challenges of the QKD methods in the view of constructing networks (the main challenges are considered above). 
Even in a network realization, QKD is a point-to-point technology. 
Speaking on the physical level, 
this means that each single photon sources should be connected to a detector by a fiber (or free space) channel as well authenticated public channel for post-processing procedures. 
Despite the fact that QKD allows obtaining of symmetric key, one can note an `asymmetry' in costs of QKD hardware. 
Typically, the cost of Bob apparatus is higher, and it requires additional infrastructure (e.g., in the case of using SSPD detectors). 
On the other hand, it is possible to use Bob apparatus in a multichannel regime.
Another limitation of QKD comes from computationally cost operation for post-processing procedures.

To overcome challenges, one can suggest the following network architecture. 
Assume that the Bob apparatus is located in a data center of the company, 
and Alice's (branches) are connected to it channel for quantum safe, hybrid, or non-quantum safe information protection applications. 
Such a scheme allows one to create the proper infrastructure for the best QKD hardware (SSPD). 
It would help to use computational resources of the company's data-center for the post-processing procedure for information reconciliation.
If for two branches $A_i$ and $A_j$ have there is direct channel for QKD to the company's data center, then effectively there is secure communication between them. 
We think such a paradigm can be easily integrated in enterprise security systems.
 
The suggested approach also allows using of QKD for information protection in distributed storage protocols. 
The protection of the transmitted data is an obvious application for quantum technologies. 
Low time of quantum state life makes it not obvious for data at rest protection. 
Protect the data at rest and data in transfer appears to be like yin and yang. 
One of the strongest ways to protect data at rest is the proactive secret sharing between several locations using the HJKY 95 protocol~\cite{Herzberg1995}. 
While there is still threat of attack in some period of time of one location, the secret data has to be re-shared. 
To achieve the maximum protection this scheme requires QKD.

Quantum computing technologies pose a significant threat to information security products based on another important emerging technology --- blockchain~\cite{Tomamichel2017}. 
Recently, a possible solution to the quantum-era blockchain challenge was suggested~\cite{Kiktenko2017}. 

If we look in the more distant future the even data in process can be processed without decryption. 
There are number of conceptual works where quantum computer can process data even without knowing what is this data about. 
The untrusted computation can make processor time a 100\% service.

\section{Conclusion and outlook}

Quantum computing is an increasingly hot area for investments both from government and large companies, e.g. Google and Microsoft. 
One can then think that post-quantum era is coming even faster than it is expected. 
Thus, the actual need in quantum-safe security mechanism at least in some applications is clear today.
To summarize, we introduced the approach for selecting proper cryptographic techniques for enterprise security systems dealing with information of varying degrees of importance. 
We proposed the approach for building networks architecture allowing efficient combining of different cryptographic techniques both for communications and distributed storage technologies. 

\section*{Acknowledgments} 

We acknowledge financial support from Ministry of Education and Science of the Russian Federation (Agreement 14.582.21.0009, ID RFMEFI58215X0009).

\end{document}